\documentclass[twocolumn,showpacs,preprintnumbers,amsmath,amssymb,amsfonts,pra]{revtex4-1}
\usepackage{graphicx,epsfig}
\usepackage{dcolumn}
\usepackage{bm}

\begin{document}

\title{Effects of pulse collisions in a multilayer system with noninstantaneous cubic nonlinearity}

\author{Denis V. Novitsky}
\email{dvnovitsky@tut.by} \affiliation{B.I. Stepanov Institute of
Physics, National Academy of Sciences of Belarus,
Nezavisimosti~Avenue~68, 220072 Minsk, Belarus}

\date{\today}

\begin{abstract}
The numerical simulations of an ultrashort pulse propagation in a
one-dimensional nonlinear photonic crystal are carried out. It is
known that the relaxation of cubic nonlinearity is the reason for
the effect of pulse self-trapping in such multilayer system. In this
paper we study further implications of this effect. It is shown that
the trapped light absorbs additional low-intensity pulses which
cannot be self-trapped \textit{per se}. On the other hand, such
low-intensity pulses are subject of the so-called induced trapping
when light gets trapped due to a collision of two such pulses. We
consider the conditions for this effect in both cases of co- and
counter-propagating pulses.
\end{abstract}


\maketitle

\section{Introduction}

It is known that nonlinear response of optical media is not
generally instantaneous and is described by a certain settling time.
This relaxation of nonlinearity can be neglected if the
characteristic time of electromagnetic field (such as a pulse
duration) is much greater than the relaxation time. However, in the
modern era of ultrashort pulses, there is growing number of
situations when this neglect cannot be justified. The standard means
to calculate the relaxation of non-resonant cubic (Kerr)
nonlinearity is the so-called Debye model \cite{Akhm},
\begin{eqnarray}
t_{nl} \frac{d \delta n}{d t}+ \delta n=n_2 I, \label{relax}
\end{eqnarray}
where $\delta n$ is the nonlinear part of refractive index, $n_2$
the Kerr nonlinear coefficient, and $t_{nl}$ the relaxation time.
The latter depends on the specific mechanism of nonlinearity. In
this paper it is assumed to be of the order of several femtoseconds.

The influence of non-instantaneousness of nonlinearity on optical
response was studied in a number of works during last several
decades. Not aiming to name all of them, we can call researches on
laser beam self-focusing \cite{Fleck69, Aleshkevich, Hanson},
filament formation \cite{Fleck72}, parametric amplification
\cite{Trillo}, pulse compression \cite{Apanasevich}, modulation
instability \cite{Shih, Velchev, Leblond, Zhang}, pulse train
generation \cite{SotoCrespo}, soliton-array generation
\cite{Cambournac}, instability of speckle patterns \cite{Skipetrov},
solitary pulse dynamics \cite{Liu, Michel}, optical switching
\cite{Armaroli, Ferreira}, etc. In this paper we deal with
ultrashort (femtosecond) pulse propagation in a one-dimensional
nonlinear photonic crystal which is a set of periodically arranged
dielectric layers. Such multilayer structure can be symbolically
designated as $(AB)^N$ ($N$ is the number of pairs of layers of $A$
and $B$ type). One of the first studies of role of nonlinearity
relaxation in such a system is the paper by Vlasov and Smirnov
\cite{Vlasov} where the problem of pulse compression was under
investigation. In recent works \cite{Novit, Novit1, Novit2, Novit3}
the effect of pulse self-trapping in the photonic crystal due to
relaxing nonlinearity was discovered and discussed. The present
paper is a continuation of those researches, so that it is worth
recalling briefly the main results obtained there.

It was shown \cite{Novit} that the pulse of high enough intensity
can be trapped inside the one-dimensional photonic crystal due to
formation of dynamical nonlinear "cavity", or "trap". This trap
appears only if both linear refractive index modulation and Kerr
nonlinearity relaxation present. The range of pulse durations and
relaxation times for such self-trapping effect to be observed in the
multilayer system of several hundreds layers was studied as well:
$t_{nl}$ varies from a fraction of a femtosecond to more than $100$
fs, and $t_p$ from about $10$ fs to several hundreds fs. This
corresponds to the fast electronic mechanism of Kerr nonlinearity.
Though we do not mean any specific materials, it is believed that
such nonlinear structures can be composed, for example, from doped
glasses with rapidly relaxing nonlinearities. This leads to the
requirement of comparatively high intensities of the pulses because
of the well-known from experiments approximate proportionality $n_2
\sim t_{nl}$ \cite{Akhm}. Nevertheless, the necessary peak
intensities of the order of a hundred of GW/cm$^2$ seems to be not
excessively high for femtosecond pulses from the viewpoint of
optical damage of the materials. Polymeric materials are worth to be
mentioned due to both high nonlinearity coefficient and fast
relaxation \cite{Meng} that make them perspective for applications
in the discussed situations. In \cite{Novit2} we studied in detail
the conditions for self-trapping in different configurations of the
structure (linear and nonlinear layers, focusing and defocusing
nonlinearities, etc.) taking into account the correlation between
the nonlinearity coefficient and the relaxation time mentioned
above. Another problem studied is the spectral transformations of
light pulses interacting with nonlinear photonic crystal in the
regime of self-trapping \cite{Novit1}. In particular, under properly
chosen conditions, it is possible to generate quasi-monochromatic
radiation or quasi-continuum covering the whole band gap. Finally,
the possibility of asymmetric light transmission due to pulse
self-trapping was analyzed recently \cite{Novit3}.

In this paper we consider the situation of not a single pulse, but
of many pulses inside the photonic crystal with relaxing
nonlinearity. We are especially interested in investigation of
interaction of probe pulses with the previously trapped radiation.
This problem is studied in \ref{probe}. Another question is
connected with possibility of light trapping due to a collision of
two low-intensity pulses which are not trapped when propagate
separately. This situation, which we call the induced trapping, is
considered in \ref{ind}.

\section{\label{probe}Trapping of probe pulses}

At first, let us state the main equations used in this paper.
Propagation of an ultrashort pulse in a one-dimensional nonlinear
photonic crystal is described by the Maxwell wave equation
\begin{eqnarray}
\frac{\partial^2 E}{\partial z^2}&-&\frac{1}{c^2} \frac{\partial^2
(n^2 E)}{\partial t^2} = 0, \label{Max}
\end{eqnarray}
where $E$ is the electric field strength, $n$ is the refractive
index which depends on light intensity $I=|E|^2$ as
\begin{eqnarray}
n(z, t)=n_0(z)+\delta n (I, t). \label{refr}
\end{eqnarray}
Here $n_0(z)$ is the linear part of the refractive index varying
periodically along the $z$ axis, and the nonlinear addition $\delta
n (I, t)$ behaves according to Eq. (\ref{relax}). Further we
consider femtosecond light pulses of Gaussian shape with the
amplitude $A=A_m \exp(-t^2/2t_p^2)$, where $t_p$ is the pulse
duration. To analyze the interaction of such a pulse with a
nonlinear photonic crystal, i.e. to solve self-consistently the
system (\ref{relax})--(\ref{refr}), we use the finite-difference
time-domain method of numerical simulations which was described in
detail in Ref. \cite{Novit}. The stability of the algorithm used is
governed by the well-known von Neumann condition $\Delta t/\Delta z
\leq 1/v=n/c$, where $\Delta t$ and $\Delta z$ are the time and
space steps, respectively. If the ratio of the steps is small enough
(taking into account possible nonlinear change of $n$), one provides
the necessary stability of calculation. The level of discretization
(size of the steps) is chosen to be optimal in respect to both
accuracy and calculation time and allows to obtain the reliable
general dependencies discussed further.

The parameters of the photonic crystal [the structure of $(AB)^N$
type] used in our calculations are as follows: the linear parts of
refractive indices of the layers $A$ and $B$ are $n_a=2$ and
$n_b=1.5$, respectively; their thicknesses $a=0.4$ and $b=0.24$
$\mu$m; the number of layers $N=200$. The pulse duration is $t_p=30$
fs, and the central wavelength of the initial pulse spectrum is
$\lambda_c=1.064$ $\mu$m if not stated otherwise. The nonlinear
coefficient of the material is defined through the nonlinear term of
the refractive index, so that $n_2 I_0=0.005$; this means that the
pulse amplitude is normalized by the value $A_0=\sqrt{I_0}$. The
relaxation time of the nonlinearity of both layers is $t_{nl}=10$
fs. We adopt these parameters in this paper though the similar
effects can be obtained even in the half-linear structure
\cite{Novit2}. It is also important to note that the parameters (in
particular, the wavelength) satisfy the requirements on the sign of
group velocity dispersion studied in Refs. \cite{Novit1, Novit2}.

\begin{figure}[t!]
\includegraphics[scale=0.9, clip=]{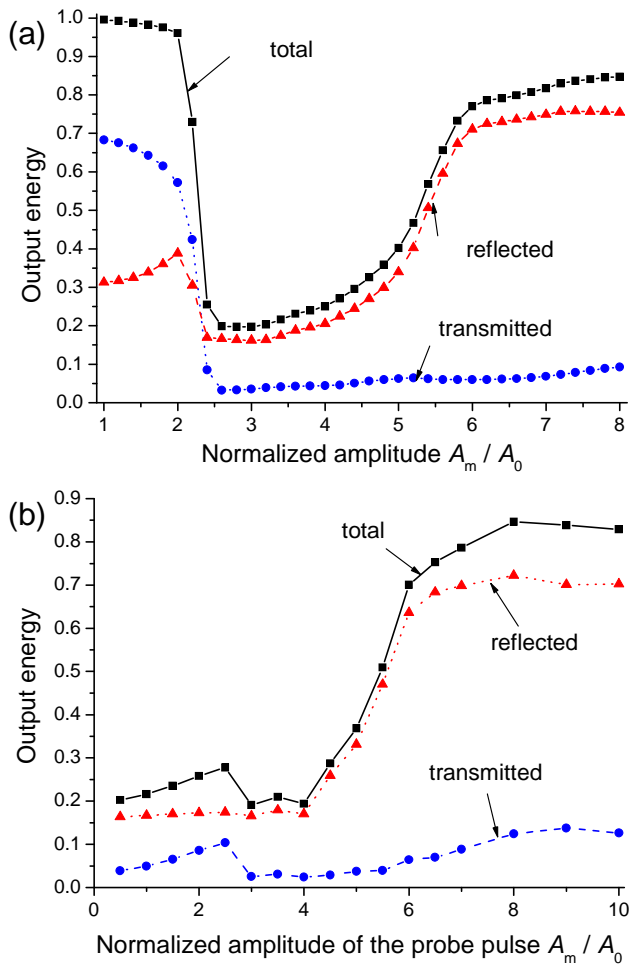}
\caption{\label{fig1} (Color online) Dependence of the output light
energy (normalized to the input energy) (a) on the peak amplitude of
the incident pulse, (b) on the peak amplitude of the probe pulse
interacting with the photonic crystal in the excited state (after
trapping of the initial pulse with the amplitude $A_m=3 A_0$). The
energy is integrated over the time $300 t_p$.}
\end{figure}

Figure \ref{fig1}(a) shows the dependence of the output energy
leaving the photonic crystal during a certain time interval after
pulse incidence on the pulse amplitude. The output energy is
calculated by integration of intensity of light leaving the photonic
crystal over time. The relative output energy normalized by the
input one is
\begin{eqnarray}
W=\frac{Q^{out}}{Q^{inc}}=\frac{\int I^{out} (t) dt}{\int I^{inc}
(t) dt}, \label{output}
\end{eqnarray}
where $Q^{inc}$ and $Q^{out}$ are the absolute values of energy of
the incident and output light, respectively; $I^{inc}$ and $I^{out}$
are the corresponding intensity profiles. The value calculated by
Eq. (\ref{output}) at the input edge (incidence plane) of the
structure gives the normalized energy of the reflected light, while
the energy calculated at the exit edge corresponds to the
transmitted light. The sum of these two values is the total output
energy. The relative output energy in Fig. \ref{fig1}(a) is
integrated over the time $300 t_p$ which is approximately ten times
larger than the pulse transmittance time in the linear regime (about
$30 t_p$). The dip seen in the curve for the total output energy is
the feature of self-trapping of the pulse. This means that the
pulses with the amplitudes in the certain range are trapped inside
the photonic crystal. Outside this range, the low-intensity pulses
transmit through the system, while the high-intensity ones are
mostly reflected.

\begin{figure}[t!]
\includegraphics[scale=0.9, clip=]{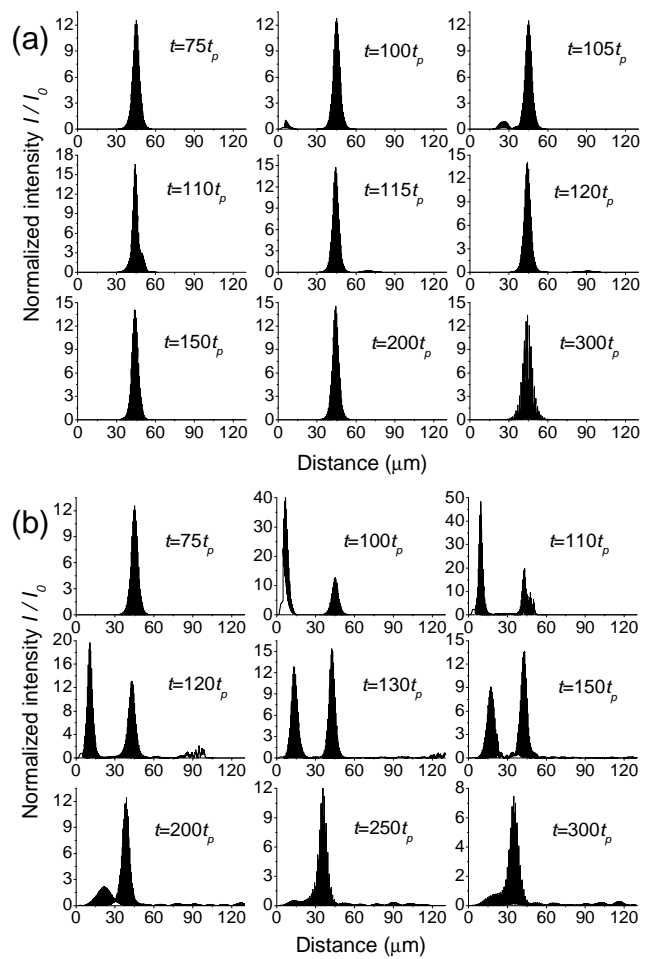}
\caption{\label{fig2} Distribution of light intensity inside the
photonic crystal at different time points. The probe pulse amplitude
is (a) $A_m=A_0$, (b) $A_m=7 A_0$. The peak amplitude of the first
pulse is $A_m=3 A_0$.}
\end{figure}

\begin{figure}[t!]
\includegraphics[scale=0.87, clip=]{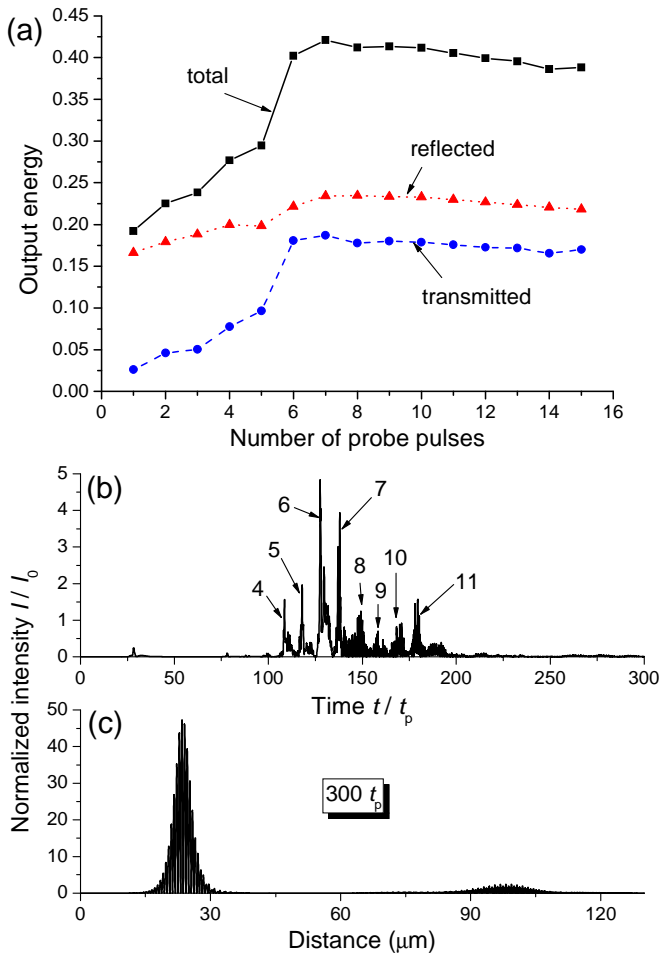}
\caption{\label{fig3} (Color online) (a) Dependence of the output
light energy (normalized to the input energy) on the number of probe
pulses. The amplitude of the probe pulses and of the initial one is
$A_m=3 A_0$; the energy is integrated over the time $300 t_p$. (b)
Transmitted intensity profile and (c) intensity distribution inside
the photonic crystal (at time $300 t_p$) for the case of $11$ probe
pulses.}
\end{figure}

\begin{figure}[t!]
\includegraphics[scale=0.9, clip=]{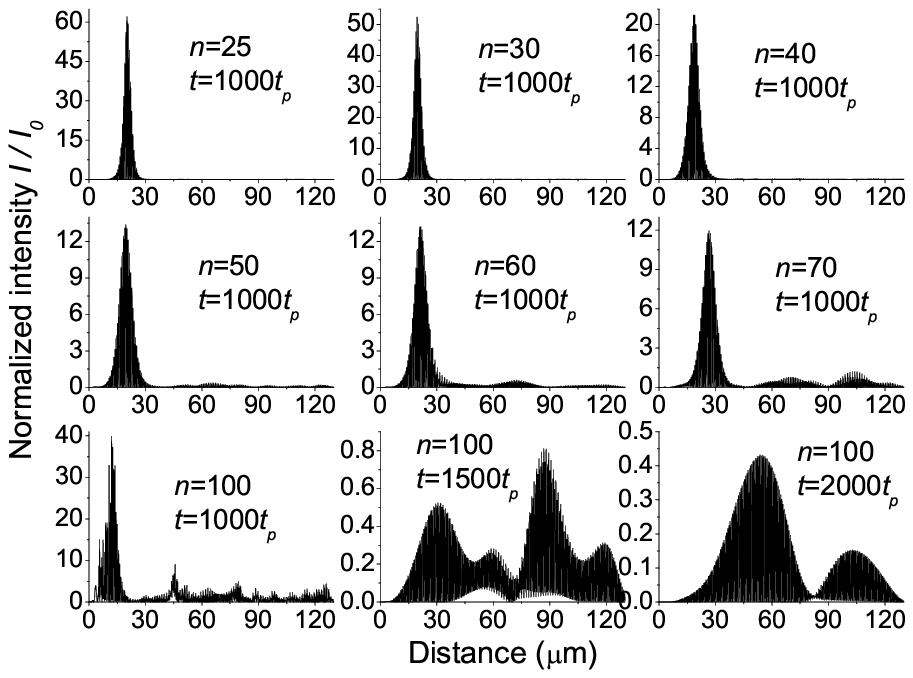}
\caption{\label{fig4} Distributions of light intensity inside the
photonic crystal for different number of probe pulses. The peak
amplitude of the initial and probe pulses is $A_m=3 A_0$. The number
of probe pulses and the time points are indicated on the panels.}
\end{figure}

The result of self-trapping is the formation of localized light
intensity distribution which is the indication of trap creation.
This trap stores most of the pulse energy for the times of the order
of several thousands $t_p$ \cite{Novit}. It is worth noting that
self-trapping does not mean light absorption, because the modulated
refractive indices remain real. Let us study what happens when the
second (probe) pulse interacts with this excited state of the
photonic crystal containing trapped light. In this case the
intensity of light in Eq. \ref{relax} is governed by the sum of the
field present at a certain space point and time instant. The second
pulse starts at the instant $100 t_p$ after the first (trapped) one.
The results of calculations of the output energy (integrated over
the time $300 t_p$) as a function of amplitude of the probe pulse is
presented in Fig. \ref{fig1}(b). It is seen that the trap formed due
to the pulse with the amplitude $A_m=3 A_0$ blocks propagation of
probe pulse with low enough intensity. This means that the energy of
the probe pulse gets stored inside the photonic crystal, so that the
trap collects more and more light. Our estimate shows that the probe
pulse with $A_m=A_0$ loses about $60 \%$ of its energy due to
interaction with the trap. The region of optimal self-trapping of
single pulses ($3 A_0<A_m<4 A_0$) is naturally the range where the
most part of both pulses is trapped. Finally, the high-intensity
probe pulse (with $A_m \gtrsim 5 A_0$) cannot be trapped. This,
however, is connected not with the breakdown of the trap but with
reflection of the high-intensity pulses seen in Fig. \ref{fig1}(a).

We prove these conclusions in Fig. \ref{fig2} where the
distributions of light intensity inside the photonic crystal at
different time instants are shown. It is seen that, by the time
$t=100 t_p$ (launch time of the probe pulse), the trap (bell-shaped
stable light distribution) formed by the first pulse with the
amplitude $A_m=3 A_0$ exists inside the multilayer system (the total
length of the system is $128$ $\mu$m). If the probe pulse has low
peak amplitude [$A_m=A_0$, see Fig. \ref{fig2}(a)], it is just
absorbed by the trap, the peak intensity of the distribution
increasing from about $12 I_0$ to $14 I_0$. This distribution is
slowly spreading with time. The situation is totally different for
the probe pulse with the amplitude $7 A_0$ [Fig. \ref{fig2}(b)]. It
is seen that this high-intensity pulse is stopped near the very
entrance of the photonic crystal and then effectively reflected.
However, some part of its light penetrates in the vicinity of the
trap and perturbs it. As a result of this perturbation, the peak
intensity of the distribution decreases (about $8 I_0$ at $300
t_p$), the trap spreads and shifts towards the entrance of the
structure. At time $t=1000 t_p$ the distribution is strongly widen
and has maximum of the order of $0.5 I_0$. In other words, the trap
looses the light energy faster than in the case of low-intensity
probe pulse and is less stable.

Nevertheless, we can state that, generally, the trap persists and
cannot be overcome by a single probe pulse. To study stability of
the trap further, we launch more probe pulses inside the photonic
crystal. The first probe pulse starts at $t=50 t_p$ after the
initial one (which forms the trap); the interval between every
subsequent probe pulses is $10 t_p$. The integral output energy as a
function of the number of probe pulses is plotted in Fig.
\ref{fig3}(a). It is seen that, at first, the part of energy leaving
the multilayer structure is growing quite rapidly. But after the
sixth probe pulse the trapped energy (in relative units) stays
approximately the same. This means that the absolute value of light
energy trapped inside the system is growing. The evidence for this
conclusion is presented in Fig. \ref{fig3}(c) where the intensity
distribution is shown for the case of $11$ probe pulses. The peak
value of intensity is about $50 I_0$ (at $300 t_p$) which is much
greater than $14 I_0$ for the single probe pulse [see Fig.
\ref{fig2}(a)]. The profile of the transmitted radiation shown in
Fig. \ref{fig3}(b) allows to identify the intensity peaks
corresponding to particular probe pulses. Notice that the first
strongly pronounced peak is connected with the fourth probe pulse,
i.e. the first, second, and third probe pulses are almost entirely
absorbed. Every following pulse appears more or less sharply in the
output of the photonic crystal.

Returning to Fig. \ref{fig3}(c), one can note not only the increase
in peak intensity of light in the trap but also the shift of the
trap towards the input of the system. Obviously, this process
continues as we launch more and more pulses in the structure. Such
situation is shown in Fig. \ref{fig4}. We also see substantial
spreading of the distribution for larger number of input pulses.
Finally, for $100$ probe pulses (lower three panels in Fig.
\ref{fig4}), the bell-shaped distribution tends to collapse sooner
than for lesser number: the trap is absent at $t=1500 t_p$ which is
about $500 t_p$ after the last pulse enters the system, while for
$25$ pulses the trap is still stable after $700 t_p$ (see left upper
panel). Nevertheless, even at $t=2000 t_p$ approximately $40 \%$ of
the energy of $100$ pulses remains inside the photonic crystal due
to multiple reflections on the numerous boundaries.

\section{\label{ind}Induced trapping}

\begin{figure}[t!]
\includegraphics[scale=0.85, clip=]{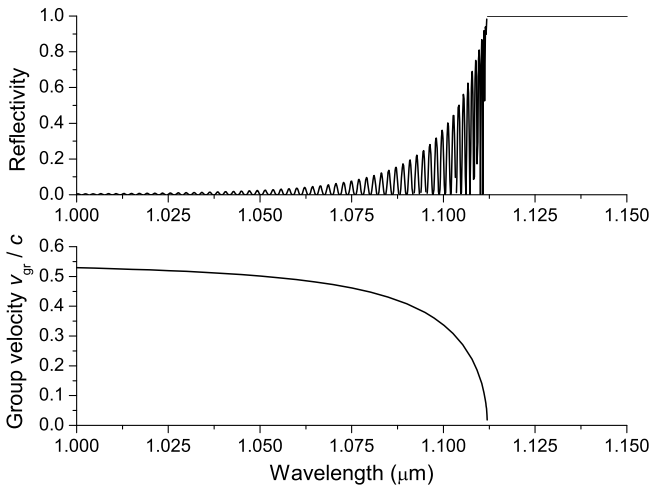}
\caption{\label{fig5} Spectral curves for reflectivity (upper panel)
and group velocity (lower panel) of the photonic crystal under
consideration. The parameters of the structure are given in the
text.}
\end{figure}

\begin{figure*}[t!]
\includegraphics[scale=0.9, clip=]{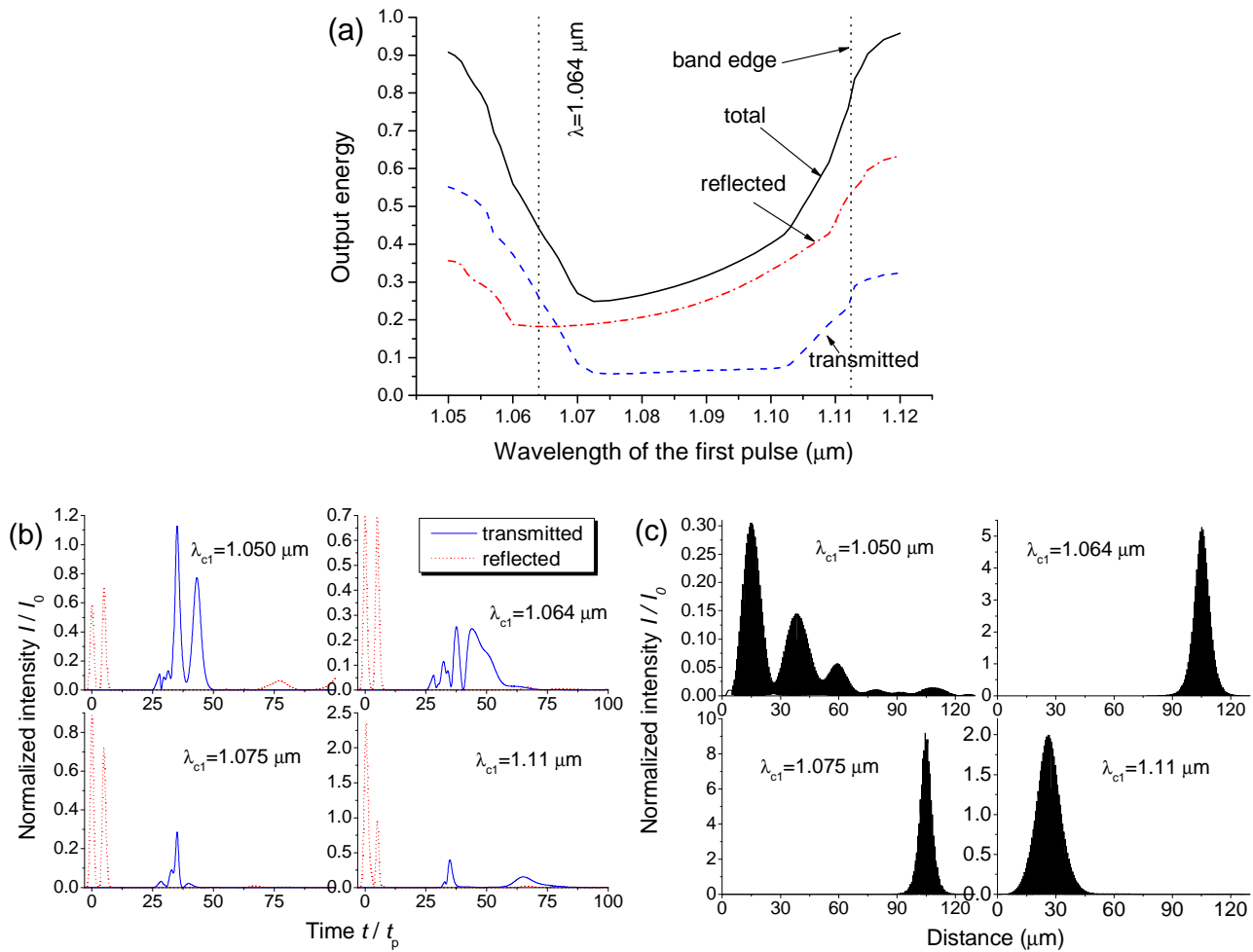}
\caption{\label{fig6} (Color online) (a) Dependence of the output
light energy (normalized to the input energy) on the wavelength of
the first pulse $\lambda_{c1}$. The amplitude of both pulses is
$A_{m1}=A_{m2}=2 A_0$; the wavelength $\lambda_{c2}=1.064$ $\mu$m;
the time-interval between the pulses $\Delta t=5 t_p$; the energy is
integrated over the time $300 t_p$. (b) Profiles of transmitted and
reflected light for different $\lambda_{c1}$. (c) Corresponding
distributions of light intensity inside the photonic crystal at
$t=300 t_p$.}
\end{figure*}

\begin{figure*}[t!]
\includegraphics[scale=0.9, clip=]{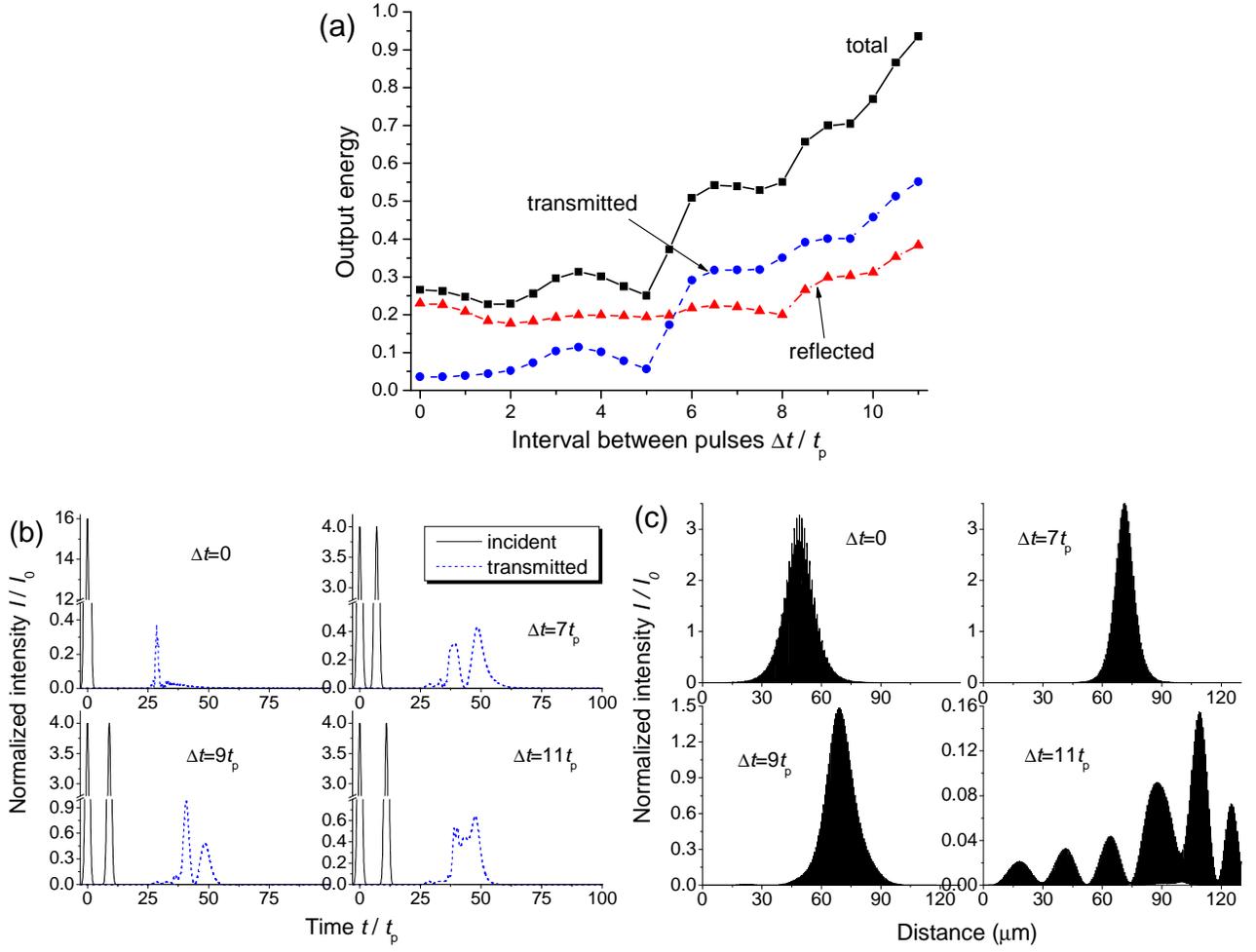}
\caption{\label{fig7} (Color online) (a) Dependence of the output
light energy (normalized to the input energy) on the time-interval
between the pulses $\Delta t$. The amplitude of both pulses is
$A_{m1}=A_{m2}=2 A_0$; the wavelengths $\lambda_{c1}=1.075$ $\mu$m,
$\lambda_{c2}=1.064$ $\mu$m; the energy is integrated over the time
$300 t_p$. (b) Profiles of transmitted and incident light for
different $\Delta t$. (c) Corresponding distributions of light
intensity inside the photonic crystal at $t=300 t_p$.}
\end{figure*}

\begin{figure*}[t!]
\includegraphics[scale=0.9, clip=]{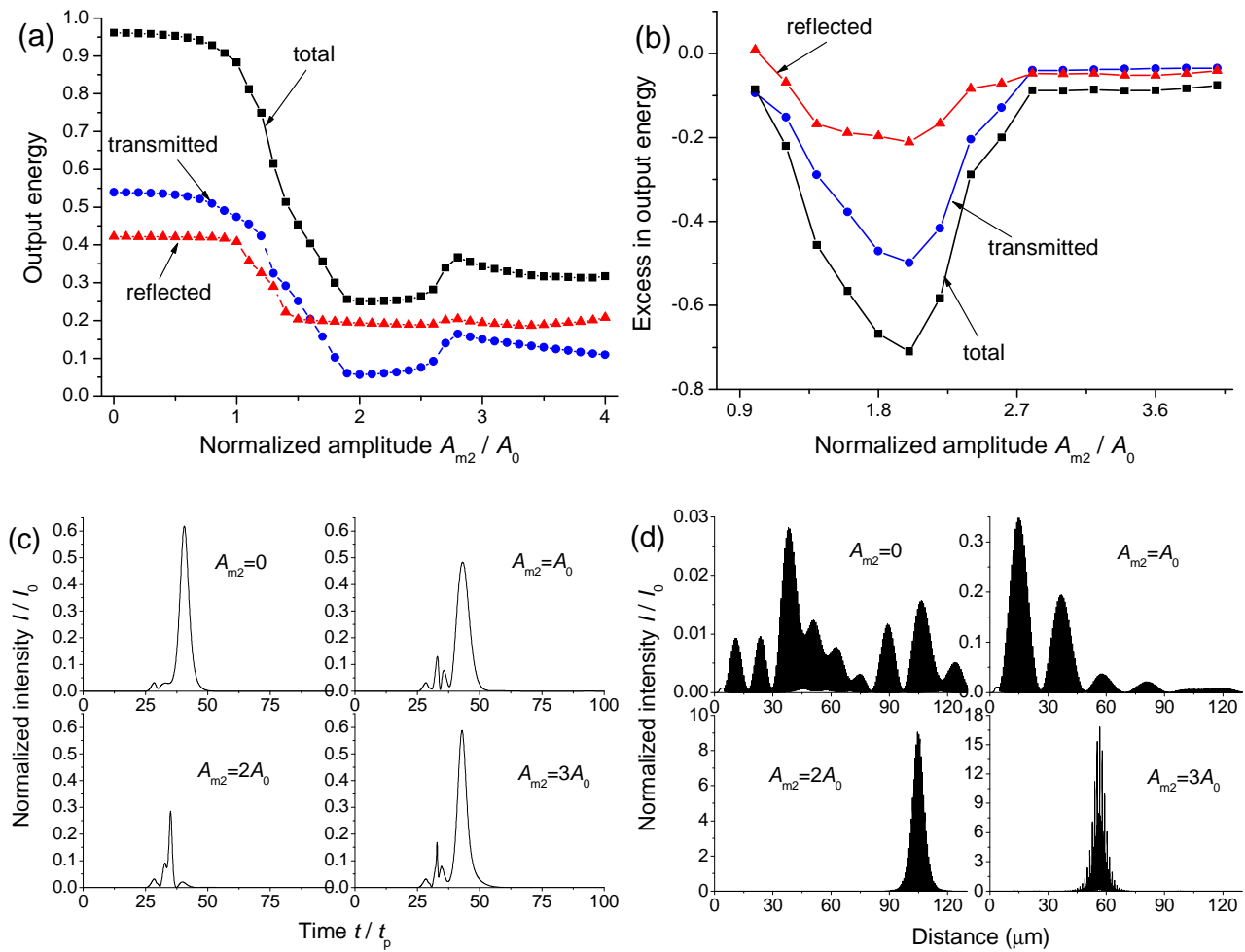}
\caption{\label{fig8} (Color online) (a) Dependence of the output
light energy (normalized to the input energy) on the amplitude of
the second pulse $A_{m2}$. The amplitude of the first pulse is
$A_{m1}=2 A_0$; the wavelengths $\lambda_{c1}=1.075$ $\mu$m,
$\lambda_{c2}=1.064$ $\mu$m; the time-interval between the pulses
$\Delta t=5 t_p$; the energy is integrated over the time $300 t_p$.
(b) The excess in output energy in comparison with the case of
single $A_{m1}$ and $A_{m2}$ pulses. (c) Profiles of transmitted
light for different $A_{m2}$. (d) Corresponding distributions of
light intensity inside the photonic crystal at $t=300 t_p$.}
\end{figure*}

\begin{figure}[t!]
\includegraphics[scale=0.9, clip=]{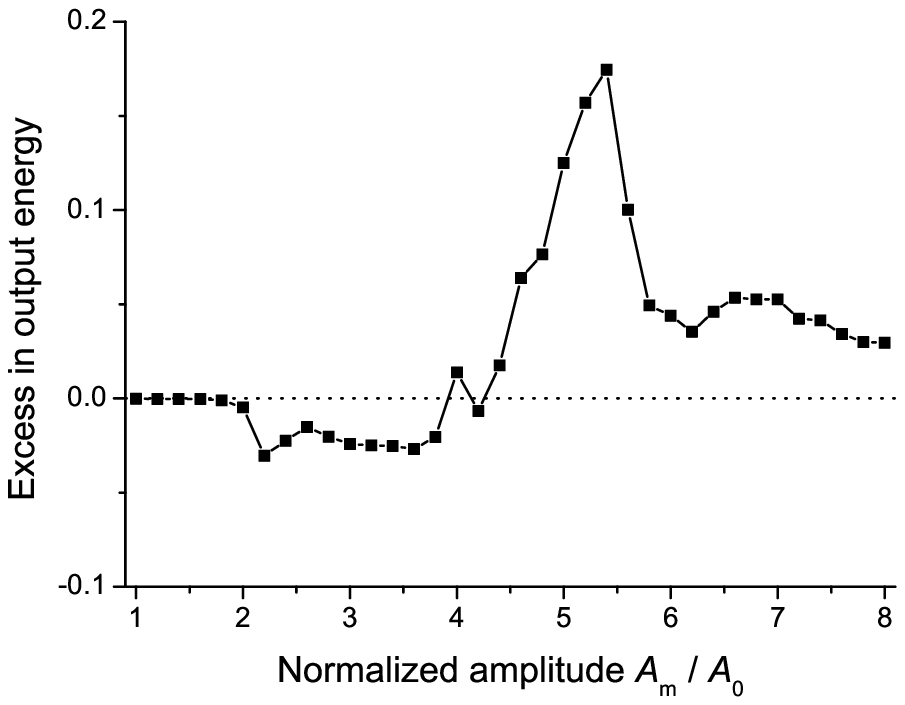}
\caption{\label{fig9} The excess in total output energy in the case
of two counter-propagating pulses in comparison with the case of a
single $A_m$ pulse. The wavelength $\lambda_c=1.064$ $\mu$m; the
energy is integrated over the time $300 t_p$; both pulses start
simultaneously.}
\end{figure}

In Fig. \ref{fig1}(a) we see that the transition between the regimes
of propagation and trapping is an abrupt one. Therefore the
amplitude of this transition which is approximately equal to $2.2
A_0$ can be called the critical amplitude. The pulses with
subcritical intensity freely propagate through the photonic crystal
(the most part of light is simply reflected and transmitted), while
the supercritical pulses are self-trapped. In this section we study
the possibility for light trapping as a result of collision of two
pulses with subcritical intensities. We refer to such a situation as
\textit{the induced trapping}.

The first possible scheme is the collision of two co-propagating
pulses. This scheme implies that the second pulse must move faster
than the first one and overtake it at a certain point of space. As a
result, the light intensity is to increase in the vicinity of this
point and formation of the trap is expected due to summation of the
fields of two subcritical pulses. To demonstrate possibility of such
course of events, we should select the proper values of the
parameters of the pulses, so that they could meet inside the
photonic crystal. It is known that the group velocity $v_{gr}$ of
the pulse in the periodic structure depends on the wavelength: it is
large in the transmission band and decreases towards the band gap
where it vanishes. The reflectivity spectrum and group velocity
behavior near the band edge of the system considered in this paper
are shown in Fig. \ref{fig5}. It is seen that the group velocity at
the central wavelength $\lambda_c=1.064$ $\mu$m is about half of the
light speed ($v_{gr}=0.5 c$). Therefore, we keep the central
wavelength of the second ("fast") pulse unchanged
($\lambda_{c2}=1.064$ $\mu$m) and vary the central wavelength
$\lambda_{c1}$ of the first ("slow") pulse.

Note that in our consideration, we neglect the dispersion of the
materials of the layers, i.e. the change of the refractive index
with frequency. This is justified by the very slow rate of this
change for many optical materials (such as glasses) in the spectral
range of our interest, namely near infrared region \cite{Palik,
Weber}. In addition, we take only narrow section of this range from
about $1.05$ to $1.12$ $\mu$m. Thus, the only source of dispersion
in the further study is the structural dispersion caused by the
order of the photonic crystal layers.

The results of calculations of the output energy as a function of
$\lambda_{c1}$ are presented in Fig. \ref{fig6}(a). The time
interval between the peaks of co-propagating pulses is $\Delta t=5
t_p$; their amplitudes are equal and sub-critical
($A_{m1}=A_{m2}=2A_0$). As it was expected, the effective trapping
occurs for $\lambda_{c1}>\lambda_{c2}$ (i.e. when
$v_{gr1}<v_{gr2}$), so that it can be unambiguously interpreted as a
result of pulse collision. The details of pulse interaction are
clarified in Figs. \ref{fig6}(b) and \ref{fig6}(c) where the
resulting intensity profiles and distributions (at $300 t_p$) are
plotted. It is seen that, when the first pulse moves faster
($\lambda_{c1}=1.05$ $\mu$m) than the second one, there are two
sharp peaks at the output. For the pulses with equal wavelengths
($\lambda_{c1}=1.064$ $\mu$m), these peaks are much less pronounced
and about a half of the energy of the pulses stays inside the
system. There is the bell-shaped light distribution near the very
exit of the structure [see Fig. \ref{fig6}(c), upper right panel]
which indicates the trap formed in this case. This trap is perhaps
due to multiple reflections from the interfaces between layers as
well as the reflection from the output boundary of the photonic
crystal. This is corroborated by the position of the trap (near the
very output) and the structure of the transmitted radiation (two
peaks of lowered intensity, in contrast to the single peak seen in
the case of optimal induced trapping at $\lambda_{c1} = 1.075$
$\mu$m). It is worth noting that the interaction between the pulses
can be observed when they move close enough one to another, i.e. the
interval between pulses is not very large.

At $\lambda_{c1}=1.075$ $\mu$m we have nearly optimal induced
trapping when there is the single low-intensity pulse at the exit.
This transmitted pulse corresponds to the first incident one, while
the second one is almost entirely absorbed. The trap forms close to
the exit and has comparatively high peak intensity (about $9 I_0$).
As we move further towards the band gap shown in Fig. \ref{fig6}(a)
by the dotted vertical line, the first pulse gets more and more
reflected, so that the transmittance starts to increase as well.
However, even near the very band edge (for $\lambda_{c1}=1.11$
$\mu$m), some part of the light energy (about $40 \%$, to be exact)
is trapped. The corresponding distribution shows that the trap forms
near the entrance of the structure. This can be interpreted as a
result of the collision of the very slowly moving first pulse (or
the rest part of it which was not reflected) and the second one. The
resulting distribution has large width and low peak intensity (about
$2 I_0$). For larger $\lambda_{c1}$, the output energy rapidly
increases to the unity as we are now inside the forbidden gap.

Thus, the effect of induced trapping can be observed in the rather
wide range of wavelengths of the pulse near the band edge of the
photonic crystal. This range is limited, at a given interval between
the pulses, by the length of the structure and, on the other hand,
by the reflectance near the band edge. The length of the system is
important at small differences between speeds of the pulses when one
of them needs much time to overtake the other; at best, they collide
near the exit of the system. The importance of the band edge appears
in the opposite case of large velocity difference when, though the
pulses collide near the entrance of the system, the most of the
"slow" pulse is reflected.

In Fig. \ref{fig7} we study dependence of the induced trapping of
two identical co-propagating pulses (with amplitudes $2 A_0$) on
interval $\Delta t$ between them. The central wavelength of the
first pulse ($\lambda_{c1}=1.075$ $\mu$m) is nearly optimal for
trapping. $\Delta t=0$ means that, in fact, we have a single pulse
with the amplitude $4 A_0$, i.e. the supercritical one. The trapping
with approximately the same efficiency is observed for $\Delta t
\leq 5 t_p$ but then the output energy starts to increase in a
stepwise manner. This is accompanied by the rise of transmitted
light as the trap forms nearer to the exit of the system (see the
panels corresponding to $\Delta t=7 t_p$ and $9 t_p$). Finally, at
$\Delta t=11 t_p$, the collision do not happen: the interval is too
large for the pulses to have time to collide. This is the same limit
due to the finiteness of the length of the photonic crystal.

The next question is the dependence of the induced trapping on the
amplitudes of the pulses at the fixed values of other parameters (we
adopt $\Delta t=5 t_p$ and $\lambda_{c1}=1.075$ $\mu$m). Without
loss in generality, we can vary only the amplitude of the second
("fast") pulse $A_{m2}$ leaving the amplitude of the first one the
same as previously ($A_{m1}=2A_0$). The results of calculations at
different $A_{m2}$ are shown in Fig. \ref{fig8}. It is seen in the
panel \ref{fig8}(a) that the output energy starts to decrease at
$A_{m2}>A_0$ and reaches the minimum at approximately $2 A_0$. At
$A_{m2}>2.5 A_0$ there are the local peaks in the curves of total
and transmitted energy. These peaks can be interpreted as a result
of effective trapping of the second, high-intensity pulse preventing
the collision. This conclusion is proved in Fig. \ref{fig8}(c): the
transmitted pulse profiles have approximately the same peak
intensity in the cases $A_{m2}=0$ (the single first pulse) and
$A_{m2}=3 A_0$ (the trapping of the second pulse), while in the
intermediate variant ($A_{m2}=2 A_0$) the transmitted pulse is
strongly suppressed due to the induced trapping. The intensity
distributions in Fig. \ref{fig8}(d) demonstrate the difference
between the trap formation due to the collision (at $A_{m2}=2 A_0$)
and single pulse absorption (at $A_{m2}=3 A_0$). Note that, in the
last case, the trap forms at larger distance than in the case of the
single $3 A_0$ pulse [see Fig. \ref{fig2} at $t=75 t_p$]. This means
that the dynamics of the second pulse in the two-pulse scheme still
depend on the first one.

Figure \ref{fig8}(a) does not allow to demarcate unambiguously the
regions of induced trapping and single pulse trapping. To carry out
this demarcation, we calculate the excess in output energy, i.e. the
difference between output energies in the cases of two-pulse and
single-pulse propagation. This value can be written as
\begin{eqnarray}
\Delta W=\frac{W_{12} Q^{inc}_{12}-W_1 Q^{inc}_1-W_2
Q^{inc}_2}{Q^{inc}_{12}}, \label{excess}
\end{eqnarray}
where the relative output energies $W_{12}$, $W_1$, and $W_2$ are
defined by Eq. (\ref{output}) in the instances of both pulses,
single first and single second pulse, respectively. $Q^{inc}_{12}$,
$Q^{inc}_1$, and $Q^{inc}_2$ are the corresponding absolute energies
of incident light (at that, $Q^{inc}_{12}=Q^{inc}_1+Q^{inc}_2$). The
value $\Delta W$ shows how much extra energy leaves the system when
both pulses are launched in comparison with the single pulse cases;
if it is negative, $\Delta W<0$, then one can say that the
additional energy is trapped inside the multilayer structure due to
the interaction between the pulses. The excess value $\Delta W$
extracted from the data in Fig. \ref{fig8}(a) is shown in the panel
(b). This figure is the evidence that, in the region
$A_0<A_{m2}<2.7A_0$, the effective trapping occurs due to the
collision of the pulses (up to $80 \%$ of the total energy can be
additionally trapped). At higher values of the amplitude,
$A_{m2}>2.7A_0$, the excess value is small and approximately
constant, $\Delta W \gtrsim -0.1$, which means that only a minor
part of light is trapped due to the interpulse interaction. This
proves our previous conclusion about practically independent
trapping of the second pulse in the high-intensity regime.

Finally, we should consider the case of counter-propagating pulses
colliding inside the photonic crystal. Obviously, it is enough to
consider the pulses with identical sub-critical intensities and
central wavelengths. We take the latter to be $\lambda_c=1.064$
$\mu$m, while the former is varied. The results of calculation of
the energy excess for this case are presented in Fig. \ref{fig9}. It
is seen that, at low intensities ($A_m<2A_0$), the excess is
negligible which means that the pulses propagate independently from
one another. At higher intensities, a small negative excess appears;
this means that only a few percent of the energy of the pulses is
trapped due to their interaction. If we increase the pulse amplitude
further ($A_m>4A_0$), $\Delta W$ becomes positive, i.e. some
additional part of energy (less than $20 \%$) is released due to
simultaneous presence of both pulses inside the structure.

Thus, there is no evidence of induced trapping in the situation of
counter-propagating pulses. One can suppose that the reason is the
short interaction time between the pulses. Indeed, in the case of
co-propagating pulses, they travel one after another exchanging
energy for a long time. On the contrary, the intersection of two
subcritical counter-propagating pulses is too short-lived to lead to
any substantial result. The calculations at different wavelength
show that slowing down of pulses does not help to improve this
situation. Only at high intensities, when the pulses are trapped
very soon after launching, the traps (which are unstable in this
case) seem to be sensitive to presence of the second pulse.

\section{\label{concl}Conclusion}

To sum it up, in this paper we have considered the interaction of
many ultrashort pulses with the photonic crystal possessing relaxing
cubic nonlinearity. As it is known from our previous investigations,
there is possibility of pulse self-trapping in such a structure.
First, we have analyzed the influence of trapped light on behavior
of additional (probe) pulses. It is shown that the trap formed by
the high-intensity (supercritical) pulse can absorb probe pulses
with low-intensity, i.e. subcritical pulses which do not demonstrate
self-trapping on their own account. This effect can serve as a
peculiar absorber for low-intensity pulses. We have also studied the
changes produced in the trap by incidence of many probe pulses.

The second topic of this paper concerns the induced trapping, i.e.
the effect of light trapping as a result of collision of two
subcritical pulses. We have shown that this phenomenon occurs in the
case of two co-propagating pulses with different velocities: when
one of them overtakes the other, their interaction leads to
effective trapping of their energy. On the contrary, the collision
of two counter-propagating pulses does not allow to observe induced
trapping.

The results presented in this paper are of general character and
therefore leave the question of their particular realization open.
Here we briefly discuss some important issues. We have used the
simple and well-known Debye model of relaxation, though the
particular nonlinear media can relax according to different laws.
This raises an interesting question of dependence of the results on
the relaxation model, though it is likely that the particular model
is not as important for the effects calculated as the presence of
relaxation \textit{per se}. The choice of model is perhaps closely
associated with the choice of relaxation time. The relaxation times
used in our research (up to $10$ fs) make another demand of
appropriate materials. Such short $t_{nl}$ are characteristic, for
example, to the electronic mechanism and its combinations with other
contributions to the Kerr nonlinearity. However, our analysis is
purely phenomenological and does not include these important
details.

Another important issue is connected with the value of light
intensities necessary to obtain the effects reported, in particular
self-trapping of the pulse. Obviously, intensities should be very
large, putting a question of optimization of the system in order to
reduce the field. According to this paper, one way to reach this is
to use induced trapping of relatively low-intensity pulses instead
of one high-intensity pulse. Another possible approach is the
adjustment of the photonic structure aimed at lowering the
requirements on the materials and pulse parameters. This work is
still to be done as well as analysis of more general situations.
Such situations include taking into account of absorption, spatial
finiteness of the laser beam (we have considered only plane wave
approximation yet), and other nonlinear contributions (for example,
light frequency conversion).

\begin{acknowledgements}
The work was supported by the Belarusian Foundation for Fundamental
Research (Grant No. F11M-008). The author is grateful to the
referees of \textit{J. Opt.} for interesting and inspiring comments.
\end{acknowledgements}

\end{document}